\documentclass[useAMS,usenatbib,referee]{mn2e}
\usepackage{graphicx}
\usepackage{epsfig}
\DeclareGraphicsExtensions{.pdf,.png,.jpg}

\title[CDM ]{ Evidences for Collisional Dark Matter In Galaxies? }

\author[Salucci]{P Salucci $^{1,2}$\thanks{E-mail:
 salucci@sissa.it} N. Turini$^{3,4}$\thanks{E-mail:
  nicola.turini@pi.infn.it}\\
$^{1}$SISSA/ISAS, International School for Advanced Studies, Via
  Bonomea 265, 34136, Trieste, Italy\\
$^{2}$INFN, Sezione di Trieste, Via Valerio 2, 34127, Trieste, Italy\\
$^{3}$University of Siena, Dipartimento DSFTA, Strada
  Laterina 2, 53100, Siena, Italy\\
$^{4}$INFN, Gruppo collegato di Siena, Via Roma 56, 53100, Siena, Italy
}

\begin{document}
 
\date{Accepted .,
  Received ...,
  in original form ... .}

\maketitle

\begin{abstract}
  
  The more we go deep into the knowledge of the dark component which embeds the stellar component of galaxies, the more  we realize the profound interconnection between them.
We show that the scaling laws among the structural properties of the dark and luminous matter in galaxies  are  too complex to derive from two inert components that just share the same gravitational field. In this paper we review the 30 years old paradigm of collisionless dark matter in galaxies.  We found that their dynamical properties show strong indications that the dark and luminous components have interacted in a more direct way over a Hubble Time. The proofs for this are the presence of central cored regions with constant DM density in which their size  is related with the disk lenghtscales.  Moreover we find that the quantity  $\rho_{DM}(r,L,R_D) \rho_\star (r,L,R_D)$  shows, in all objects, peculiarities very hardly explained in a collisionless DM scenario. 
\end{abstract}

\section{Introduction}
 
The mass distribution in  Spirals  (and in any other  galaxy) is largely dominated by a dark component. This  comes from their kinematics and from  weak and strong  lensing effects that  arise only in  gravitational potentials dominated by such a component (\cite{rub,bos,Sc}).  Moreover,  the analysis of the  CMB fluctuations spectrum  and a number of cosmological measurements unavoidably  point to a  scenario in which a  Dark Massive Particle  is  the responsible for the  mass discrepancy  phenomenon in Galaxies  and Clusters of Galaxies ( \cite{pl1}). 
Alternative scenarios to the Dark Matter do exist (e.g.\cite{mo}), but, in the light of the  evidences  reported above  and of their inability to address the crucial issue of how galaxies did form,  they are  far less convincing than the DMP scenario.  We  associate, as usual,  the huge {\it local} mass discrepancy in galaxies with the presence of  surrounding halos  made  by  a massive elementary  particle that lays outside  the HEP  Standard Model (e. g.: see \cite{Ber}).  This particle also does not interact significantly with atoms, photons and with itself,  through  strong, weak and electromagnetic force.  This  does not strictly require that  DMP must interact with the rest of the Universe only through  gravitational force,  but  that, such eventual  interaction must  be  much  weaker with  respect to the  ordinary baryonic  matter vs baryonic  matter interaction. Moreover, no current observation prevents the existence of interactions between the dark and  the luminous sector of elementary particles that result relevant in the  galaxy formation context.  
However, so far, the simplest  dark matter scenario has been routinely  adopted, according to which the  DM halos  are  made by WIMP particles,  more precisely by collisionless cold dark  massive particles that interact very feebly with themselves and atoms.  These particles are thought to  emerge in SuperSymmetric extensions of the Standard Model of Elementary Particles (e.g.  \cite{Ber}).

Although large scale observations are in agreement with the predictions of  this scenario,  recently serious  reservations are  mounting against it.  In fact, at the  galactic scale  masses of $M<10^{11-12} M_\odot$,  the predicted WIMP/$\Lambda$CDM  dark matter halos are much more numerous than those detected  and show very different structural properties with respect  to those inferred by the  internal motions of galaxies (e.g. see \cite{dmaw}). The questioning issues for  the  WIMP particle  are  well known  as the ``missing satellites'' ( \cite{kl}),  the  ``too big too fail'' (\cite{boylan})  and the lack  of a  cuspy central density profiles in the  DM  halos (\cite{gen,Sp,oh} and reference therein). There are proposals in which astrophysical processes could modify  the predictions of the N-body  $\Lambda$CDM  models and the related density profiles  to fit the observations (e.g. \cite{vogel, pontzen, dicintio,rac}).  However,  this modelizations are growing in number and in diversity (\cite{ks}) and the cores formation via hypothetical strong  baryonic feedbacks  requires ad hoc fine tuning. Let us also  remind that WIMP particles  have not convincingly  been detected in underground experiments (see e.g. \cite{Fr})  and they  have not  emerged even in the most energetic LHC proton-proton collisions (e.g.\cite{cms}). Finally, the  X and gamma ray radiation coming from  annihilating WIMP particles at the center of our and nearby galaxies has not unambiguously been detected ( \cite{Fr}, e.g. \cite{FL, L+}).  Thus, to claim that  $\Lambda$CDM is not anymore the  forefront cosmological scenario for dark matter will bring no surprise.

Recent alternative  scenarios for dark matter  point  to  a sort of significant self-interactions between the dark particles which seems suitable to explain the observational evidence which has created the $\Lambda$CDM crisis.  Among those,  the Warm Dark Matter,  the axion as a Bose-Einstein condensate  and the self-interacting massive particles scenario (e.g. \cite{Fr, KD, srm,dvss}) are the most promising.  Their common  characteristic is that, at galactic scales,  dark matter  stops to be collisionless  and it starts to behave  in a way  which could make it compatible with observations.
However,  also these scenarios  hardly explain  the fact that we  continue to find  that in galaxies,  dark and luminous matter are  extremely well  correlated (e.g.  \cite{ggn}).  Thus, we have to envisage the possibility of a direct interaction between the  dark  particles and the galaxy atoms and photons leading  to major cosmological/astrophysical consequences. In fact,  the dark-luminous coupling that emerge in spirals is so intricate  that it is  extremely difficult  to frame it in a scenario in which the dark and the luminous galactic components are completely separated but through their gravitational interaction.

In this paper, just by varying the I magnitude,   we will investigate the whole  family of {\it normal} spirals, i.e.  all disk systems of Sb-Im Hubble types and with  I-magnitudes in the range  $-17.5\leq  M_I\leq -24$,  whose  corresponding 1)  halo masses range between $10^{11}  \ M_\odot$  and  $10^{13}  \ M_\odot$, 2)  disk masses  between $10^{9}  \ M_\odot$  and  $10^{12}  \ M_\odot$, 3)  optical radii between $3$ kpc and $30$ kpc  and  4) optical velocities between  $80$ km/s and $300$ km/s.

In detail, the matter in these  galaxies  has two different components: a luminous one, with its sub components (stellar disk, stellar  bulge, HI disk)  proportional to the corresponding luminosity densities, and a dark one, which results  distributed in a very different way.  In detail, the HI disk is somewhat dynamically important in regions well outside those considered in this paper and,  by selection,  the galaxies we consider here have a negligible bulge.

The stars in late type spiral galaxies are the main baryonic component in the inner regions and are settled in thin disks with an exponential surface density distribution (\cite{Freeman})
 \begin{equation}
\mu (r) = \Sigma_0 e^{-r/R_D}\,, \quad 
\end{equation}
where  $\Sigma_0= (M_d/L) I_0$  is the central surface mass
density, with $I_0$ the central luminosity density,   $L $ is the total
luminosity in the I band.
The uncertainties  in the measurement of the lenghtscales   $R_D$ are reasonably between 5\% and  10\%

Given the aim of this work $\rho_\star (r,L,R_D)$ is derived by assuming  the 3D geometry of  the spiral disks  as cylinders with circles of radius $3\  R_D$ as bases  and, inspired by  spirals of our Local Group,  $0.1 R_D$ high on the rotational plane.  We also assume no dependence  of the stellar density with the  $z$  cylindrical coordinate  Then: 
  \begin{equation}
 \rho_\star(r)=\mu(r,L)/(0.1 R_D)
 \end{equation}

 We have also  considered  other reasonable  modelization for the 3D stellar distribution (see  Appendix A),  but the  results found in this paper  are independent of this choice.

A detailed investigation of the inner combined  kinematics of thousands of spirals ( \cite{pss} hereafter PSS, \cite {ys,cat}) allows us to determine the  distributions of their dark and luminous components which show an  universal behavior,  namely,  specific  functions of  1) the disk mass $M_D$, (or equivalently  of the  disk Luminosity or  of  the  halo mass) and of 2) the disk size $R_{opt}\equiv 3.2 \  R_D$ ( \cite{ps91, pss, ks}). This picture was confirmed also by the mass decomposition  of hundredths individual RCs (e.g.  \cite{Sp, kf} and it will allow us to investigate the coupling of  dark and luminous matter in spirals of all luminosities 

Let us notice  that in this  paper  the DM halo density has a cored inner distribution, rather than the cuspy NFW profile. This originates by the results in \cite{pss, ys} but is  is also further  well justified in literature
\cite{ dB} and  it will not be discussed here.  

The outcome of this study  call for a new dark sector as a  portal for a  direct DM/LM interaction capable to modify the galaxy mass distribution on a  Hubble time scale.

In the second section we will present  the evidences  of  the  tight dark-luminous coupling  in spirals and  the resulting  scaling laws of the structural parameters of  the luminous and dark matter that call for a collisional nature of the  dark matter particle. In the third section we  show  the  direct imprint of  such a process in spirals. 

In the next section we start to  investigate its underlying physics. 

Radii are in kpc, velocities in km/s

\section{The  Coupling of  Dark and Luminous matter in Spirals }

In this section we will present new and old evidence for the existence of a peculiar coupling between the dark and luminous components of spirals.
Let us recall that,  in this work  the DM density distribution  is  represented  by the well known Burkert profile,  that very successfully performs in fitting the RCs of spirals (\cite{SB, s7,bu})
 
 \begin{equation}
\rho_{DM}(r)=\frac{\rho_0 r_0^3}{(r+r_0)(r^2+r_0^2)}
 \end{equation}
 The  core radius  $r_0$ is a fundamental DM structural quantity which   defines the inner halo dark region of almost constant density $\rho_0$  

The analysis  of the coadded (and of  the individual) kinematics of a large number of spirals shows that the distributions of  dark and luminous matter are very   well represented by the above discussed URC mass model (see fig 2 of PSS),  more precisely by  the  Eqs (4)-(8) of \cite{s7}). This set of equations provides,    for the entire family  of  spirals: $\rho_{DM}(r,M_I R_D)$ and $\rho_\star (r, M_I, R_D)$,  i.e. the  dark matter and stellar density profiles  as function of their I-Band luminosity and disk length scale.
 
In order to estimate the error budget in  the determinations of the densities we should recall that:  1) the cosmic variance of the spirals mass distributions  is quite small \cite{ys})  while   2) the best-fitting 1$\sigma$ uncertainties of the three free parameters of the URC velocity model are (see Appendix B and   \cite{pss,s7} ):

\begin{equation} 
 \Delta \rho_0/ \rho_0 = 0.2,  \Delta r_0/r_0 = 0.2, \Delta M_d/M_d =0.15 
\end{equation} 
 
 Noticeably,  these fractional  uncertainties  are small and  independent of the halo mass,  so that they  affect the log-log scaling laws in  Eqs (4)-(8) of \cite{s7}) only by  small random values.

We find that the core radius  $r_0$ tightly correlates with  $R_D$, the  stellar disk lenght-scale (see Fig. 1), data are from \cite{pss} and  \cite{ks}):

\begin{equation}
\log \  r_0= 1.38\  \log  \ R_D + 0.47
\end{equation}

\begin{figure}
\center\includegraphics[width=11.5cm]{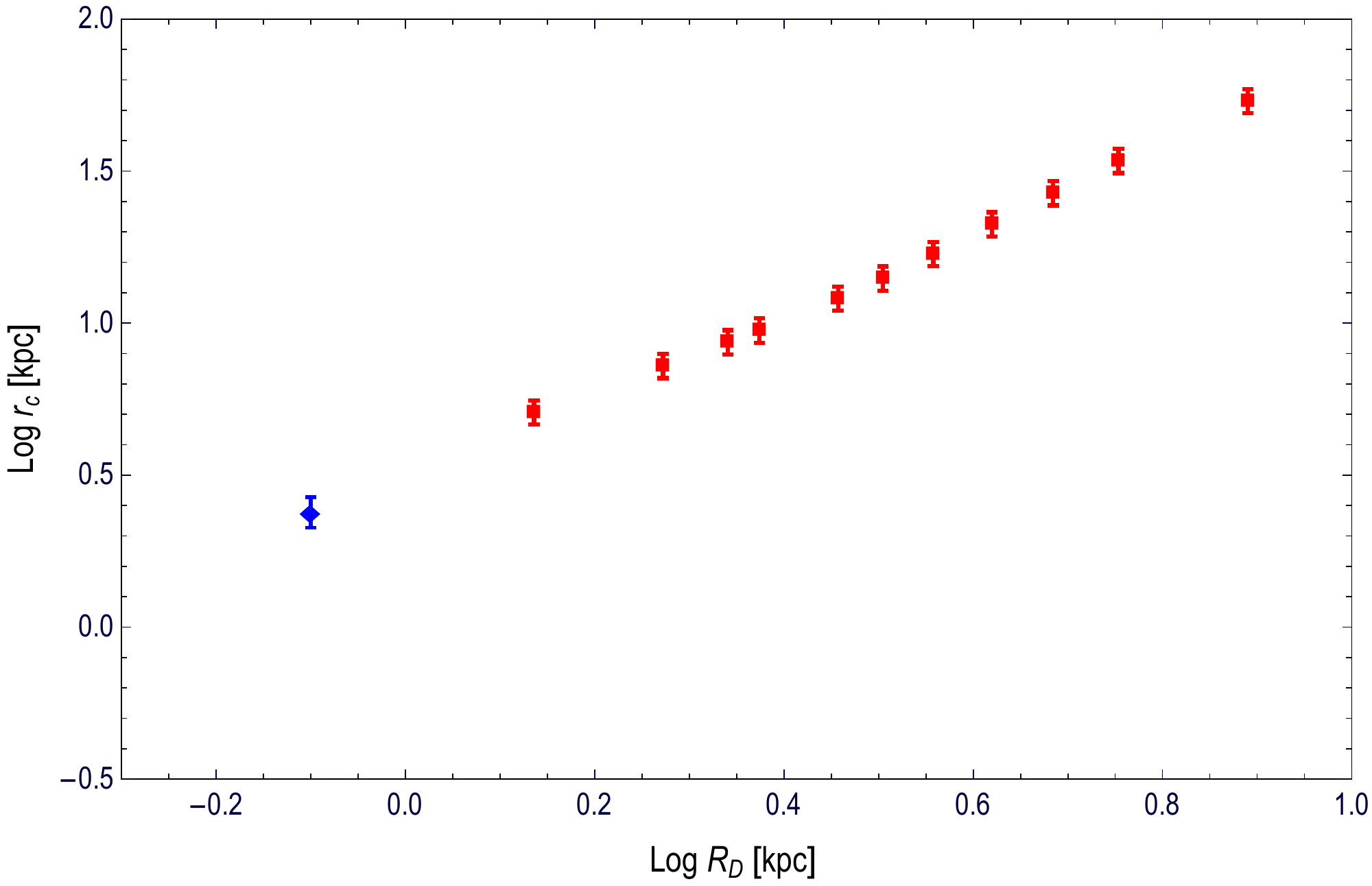}
\caption{ The $r_0$ vs $R_D$ correlation in normal ({\it red}) and  dwarf spirals ({\it blue})}.
\end{figure}

The error budget of eq (5) is enough low to make it statistically relevant. In fact, first let us notice that  the quantities involved ($r_0$ and $R_D$)  are derived  in totally  independent ways, moreover  a) the measurement errors in  $R_D$ are negligible in that  we use coadded values and   b)  the range of variation  of $r_0$ among Spirals is about  1.1 dex,  while its fitting  uncertainty is only  0.04 dex. 

This correlation is  confirmed by  \cite{D, kf, Sp} and it  extends  over several orders of magnitudes and to different Hubble Types  as  dwarfs spiral  \cite{ks}  and ellipticals \cite{mem}. The scaling law of eq (5)  has no  straightforward explanation.   In fact, while  the  origin and the values of the disk lenght-scales $R_D$ can be   traced  to the  angular momentum per mass unit  owed by their   HI proto- components  (\cite{MMW}),  the halo core radii   have certainly a different (and yet unknown) origin.  Eq (5)  is a tight    relationship   between two  structural quantities  of  spirals that, in collision less DM scenario,  are thought to arise  from totally  different physical processes.

The analysis of the URC brings up  a second intriguing coupling  between  the dark and luminous matter in Spirals:   
(\cite{s7}): 
 
\begin{equation}
\log {\rho_0\over \mathrm{g}~\mathrm{cm}^{-3}} = - (23.5 \pm 0.2) – (0.96  \pm 0.1),
\left({M_D}\over 10^{11}\, \rm M_{\odot}\right)^{(0.3 \pm 0.03)}~ 
\end{equation}

Noticeably, a similar result is obtained  in  individual objects \cite{kf}).  Eq (6) is  statistically  relevant: the relationship is monotonically decreasing and,  in spirals,  the ranges of  (log $M_D$,  log $\rho_0$) are (2.9, 1.6) \ dex, while  their  fitting uncertainties are only (0.2,0.2) dex.   Within the collisionless DM halos  scenario,  the  standard process of  the  formation of the  spiral disks  within DM halos, unlikely,  would make  the stellar disk masses and the DM halos central densities  partners in any  tight relationship.   

Furthermore, let us recall  that spirals show other intriguing relationships with no straightforward  explanation in the scenario  in which  the DM halos are made by  collisionless dark particles  1) at $r_0$, the baryonic component of the  acceleration   relates with the galaxy  luminosity  (see \cite{ggn} for details)  and 2) in dwarf Spirals  the concentrations of  dark and  the luminous matter are directly correlated (see \cite{ks} for details).  

Then,  the  structural properties of the spiral stellar disks correlate with those of the surrounding  DM halos in ways that, within  the collisionless WIMP scenario,  have not a clear physical justification.  In fact, the direct  interactions  between dark and luminous matter are absent. The indirect ones, instead, to reproduce the  observational evidence  require that  the baryons largely modify, with fine tuned models,  the total gravitational field that,  back reacting,  act in a very precise  way on the dark matter distribution  (e.g. \cite{dicintio}).

In this work,  we take the view that   the  straightforward relationships  presented  in this section  are beacon of a different scenario.

\section{A different Paradigm}

 We have seen in the previous section that in galaxies, physical  quantities,  deep  rooted in  the Dark World,  correlate  with  the most important quantities of the Luminous  World. 
Let us stress that  this  lack of a direct  explanation emerges when  we strictly follow  the view according to which  dark and luminous matter interact only through the gravitational force. Everything changes if we consider the possibility for which halo dark particles, over the  Hubble time, exchange a fraction of their  kinetic energy with ordinary matter. This opens a talking line between dark and luminous matter which can have  a dynamical role in  the  galaxy formation process, and it explains  the  above relations as a straight dynamical outcome of the above new interactions.

We, therefore,  propose a different paradigm which abandons  the assumption that the  interactions between DM and LM are only of  gravitational  nature,  and we postulate the existence of  a collisional Dark Matter Particle. Namely, a particle that directly  interacts,  in a  Hubble Time, with the ordinary  matter in an astrophysically relevant way.

Let us  stress that such an idea is not new, in fact,  it has has  proposed and discussed in the literature in a number of works \cite{ho} What it is  absolutely new here  is the observational support that  we  claim we provide at support.  

This interaction, in competition with the gravitational force that acts on the much shorter time scale  of the galaxy free-fall time, is able to modify  the  DM halos  distribution around  galaxies, namely in their central regions. One of the many possible schemes that ensues  this,  depicts that  the  dark particle interacts with components of ordinary matter and then it decays into light, hot, high momentum products that  escape from the gravitational field of galaxies. Such type of particle, evidently, does not exist in the Standard Model, neither in its  most popular extensions such SUSY.  So, in view of the above, we are thinking on a (new) dark Sector interacting with ordinary particles such as photons,  nuclei and electrons in the galaxies.  We expect, in this case,  that the DM mass should be large enough $\sim \  10^2\  GeV$  to  have easily escaped detection until now.

Respect to  previous works,   we are not postulating a specific new  type of particle and checking it  with crucial observations, but  we are instead claiming that the latter require that the DM particle should have certain exotic properties.

Without entering a very complex issue we have  to stress that  such  particle  interaction with the ordinary matter  can  be  detected.  In fact it could produce excesses in cosmic positrons or/and antiprotons  and then  be a component  of  the PAMELA and AMS detected excess. In addition, it could produce  diffuse VHE photons.  Moreover, this collisional particle, if  created  in accelerators, could be detected as missing momentum or missing masses in the particle flow of the interaction.  It is worth to notice  that  LEP data doesn't show such signature, while hadronic colliders data are more difficult to be interpreted.  However,  more powerful missing masses searches and event symmetry studies are currently performed at LHC and  there is a chance that these particles, if  produced, will be detected. Data mining of old experiments could indeed even give us  useful hints on the issue. Moreover, non accelerator searches  for DM, eg, LUX and Xenon1T,  that play a decisive role for a WIMP searches,  are { \it  currently} not tuned for such detection.
Furthermore, after the  lack of detection of  ``favorite'' SUSY particles, this new scenario opens  up a huge  field of possibilities  for  the actual nature of this collisional particles, whose precise  individuation  is beyond the aim of this work. 

Finally, although we are unable to imagine it now, we cannot  exclude  the existence of  some weird  scheme of transferring energy from baryons to WIMP particles, whose effect on the above scaling laws  would mimic the presence of  collisional interactions. However, as the  information on the DM distributions gathers, also new evidences  for a direct coupling with the baryonic components do.  
  
\section {Evidences  of an interaction between dark particles  and  atoms}

In section (2) we have presented a number of relationships that  provide motivations  for abandoning the  framework of collisionless Dark matter halos.    In this  section,   we will   investigate  other special properties and relationships of Spirals that will direct us towards  different framework  featuring the existence of  {\it direct} interactions between the dark and luminous components.   

Let us assume  spherical symmetry and introduce the DM particle pseudo pressure 

\begin{equation}
P(r)= 1/3 \  \rho_{DM}\  V(r)^2
\end{equation}

 \begin{figure}
\center\includegraphics[width=11.5cm]{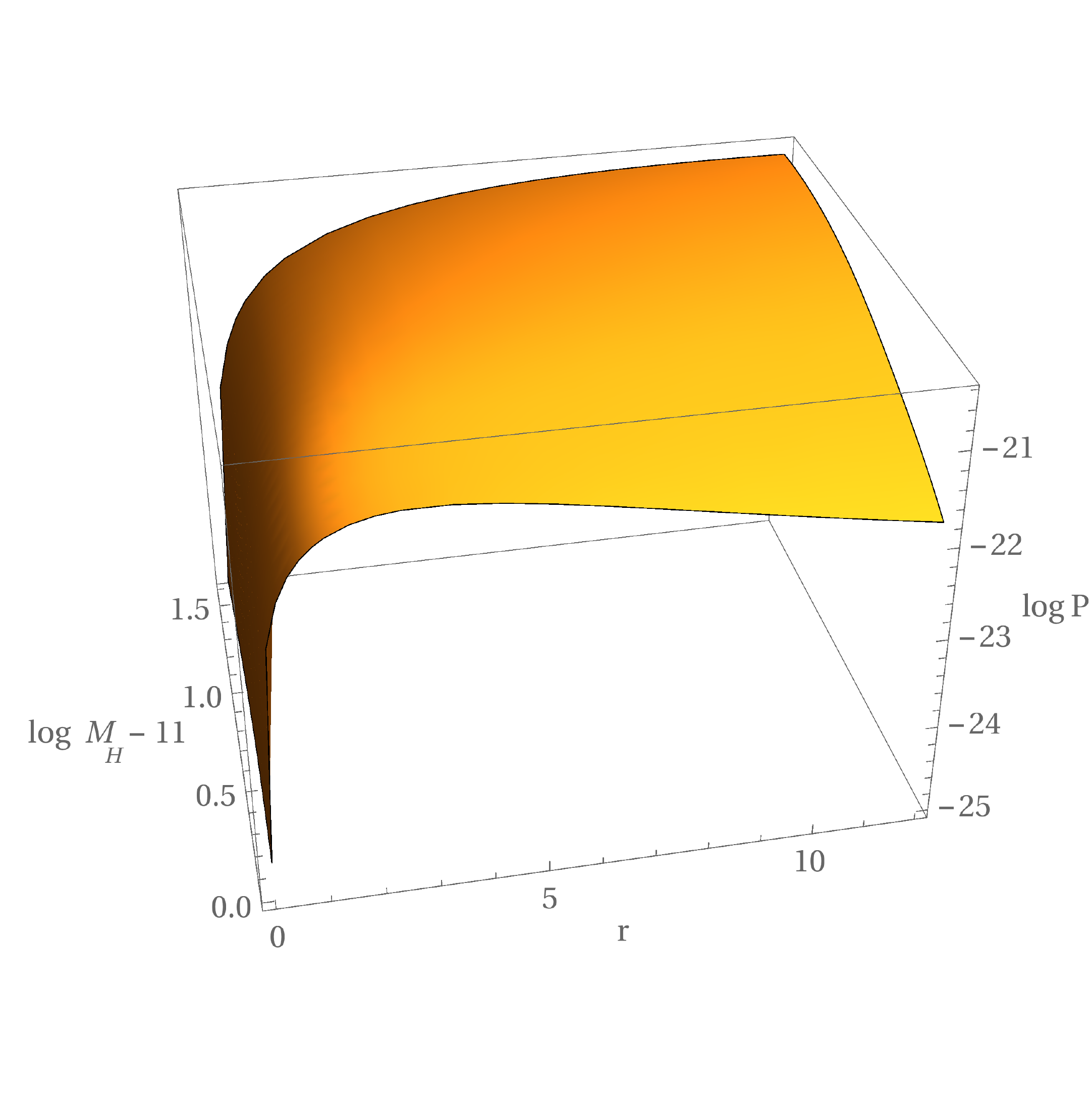}
\vskip -1.2truecm
\caption{ The pseudo pressure in Spirals  as function of halo mass (in solar masses) and radius (in kpc). The uncertainties on $log P(r)$,  propagated from those of  mass model fitting,  are 0.2 dex  }
\end{figure}

whose radial variation balances, on a dark particle at  radius $r$,  the gravitational attraction of all the matter inside $r$.  For   halos around spirals we derive this quantity from the mass modelling of the Universal rotation curve which,  for any Spiral  of halo mass $M_H$,  endows us with the corresponding values of $\rho_0$, $r_0$, $M_D$, $R_D$ and their uncertainties (see section 3).  Notice that due to the term   $V(r)^2=V^2_{DM}+V^2_{b}$  inside the pseudo pressure, $P(r)$  is a hybrid dark-luminous matter quantity. It behaves as it follows (see Fig. 2)  Its value  is zero at the center,   it rises out to  the radius  $R_{cp}$ and  it slowly  declines outward.  Thus, in spirals,  the length scale $R_{cp}$ emerges.  We  derive this quantity  by computing  the radius at which   $dP(r)/dr=0$.  $R_{cp}$ is found to tightly correlate  with the galaxy halo mass (see Fig. 3).

$P(R_{cp})$  varies less than a factor 1.5  in all galaxies,  that  might also suggests  that the  DM particles have been subjected to  some form of direct interaction with baryon.  Looking at the Fig (2) the radius  $R_{cp}$ can be seen as that at which there is a  dynamical equilibrium between the number of DM particles that, inside  this radius, get destroyed during a Hubble Time and the number of the  ones  that,  in the same period,  enter  into such a volume,  driven by  the  gravitational unbalance inside it. Notice that $R_{cp}$ is finite since the rate of  DM particle destruction by interacting with baryons, for $r > R_{cp}$, strongly declines with radius because the latter do so. Therefore, we envisage that,  in each galaxy,  inside a  volume of radius  $~R_{cp}$, the  DM density  has dynamically evolved in a  Hubble time, leading to   $ T_H d\rho_{DM}/dt/ \rho_{DM}(10^{10} y)  \simeq 1 $. Notice  that the zero-pressure gradient  line above,  if not due to the proposed radial time variation of the number of particles, has to be originated by a huge radial variation of their anisotropy, which we have  no explanation of.   

$R_{cp}$ tightly relates  also with the core radius $r_0$  (see Fig. (4)).  This may suggest   that  the physics behind the former quantity is connected with the existence of  cored  DM cored distributions. In any case, this  relationship  is not explained by the  collisionless DM models,  in which  the two  correlating quantities do not even  exist.
 \begin{figure}
\center\includegraphics[width=9.5cm]{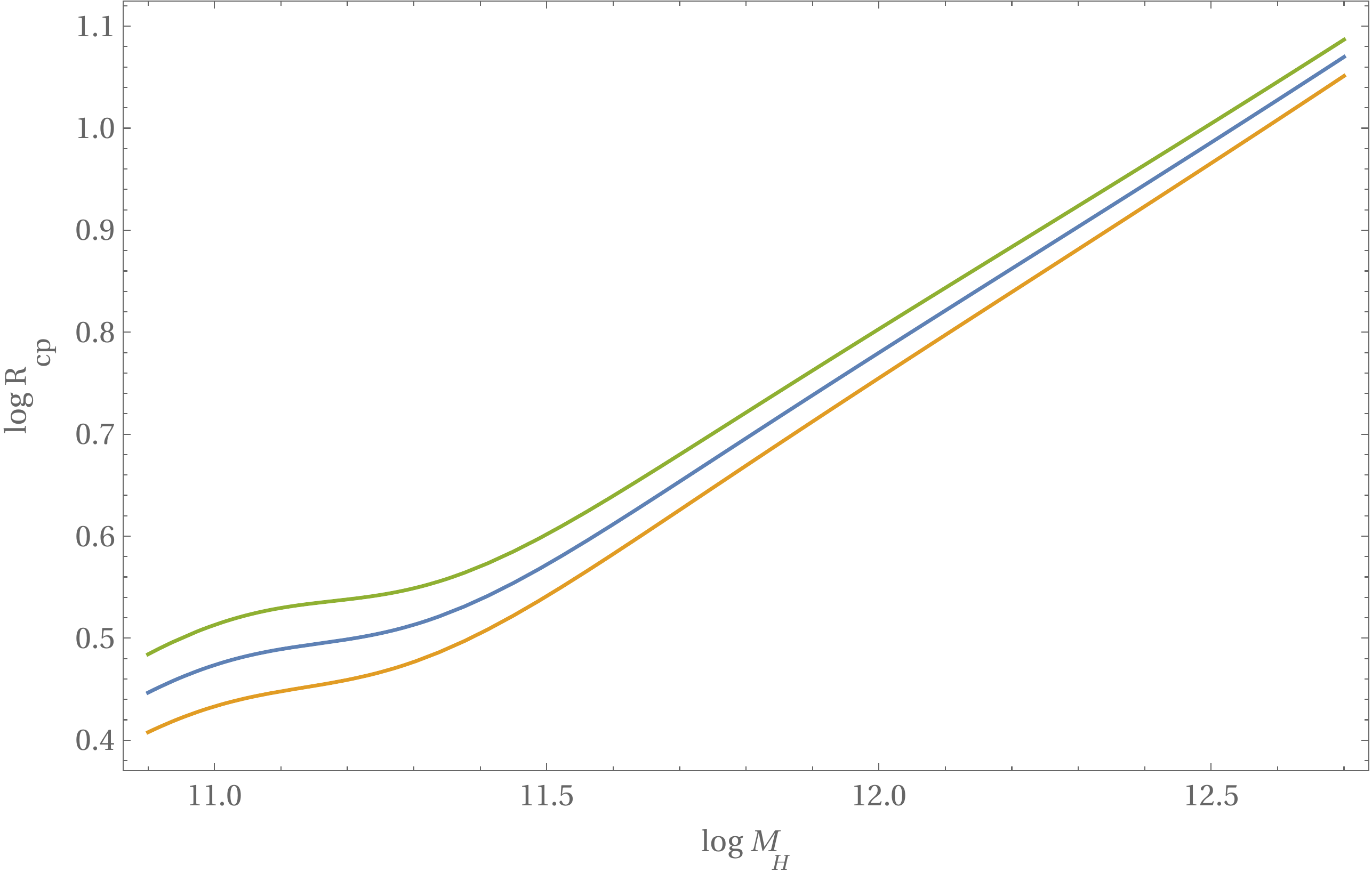}
\caption{ The relationship between $R_{cp}$ and $M_H$ (blue) . The red lines are its   2 $\sigma$  uncertainty  countour  }
\end{figure}

As in   the  self-annihilating dark matter case,   in which the well known  kernel of the astrophysical term is $K_{SA}= \rho_{DM}^2(r) $,   in  the present collisional dark matter  case, we have that

\begin{equation}
 K_C(r)=\rho_{DM}(r) \rho_\star(r)
\end{equation}

 is the kernel of the (collisional) astrophysical  term    of the    proposed DM -baryons interaction. Notice that the term  $K_C$  can consider also   the interaction between dark particles  and the  photons and neutrinos emitted by stars and supernovae, whose numbers density is, on large scales,   proportional to  $\rho_\star(r)$. 
 
\begin{figure}
\center\includegraphics[width=9.5cm]{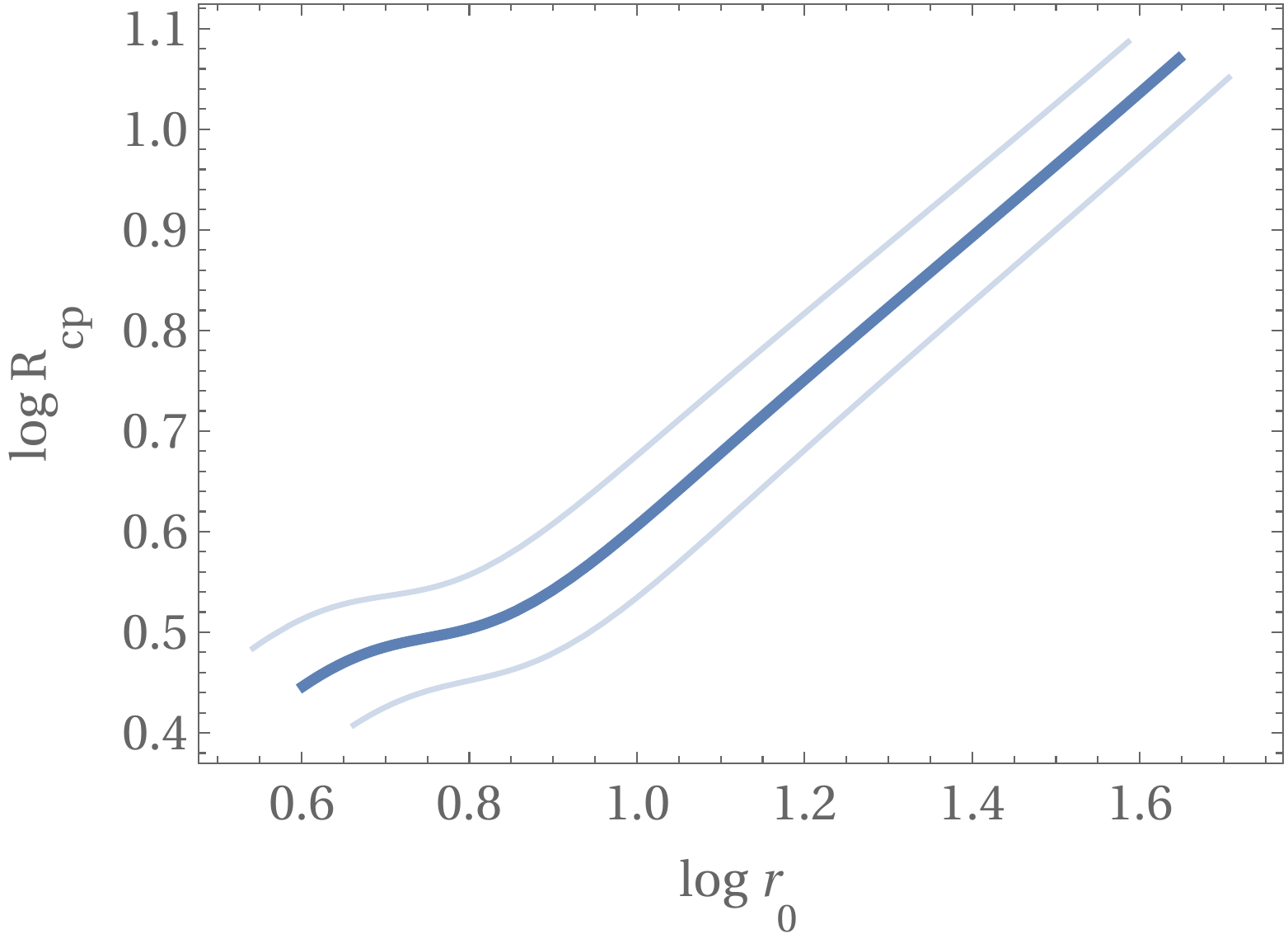}
\caption{ The  relationship between $R_{cp}$ and $r_0$  in spirals (thick line). The thinner lines are the 2 $\sigma$ uncertainty  countour }
\end{figure}

Finally, in the case of  collisionless dark matter,  $K_C(r) $ has little physical sense,  the quantity that matters, in this scenario, is instead  the sum of the two densities: $\rho_{DM}(r) +\rho_\star(r)$  that get  related via the gravitational potential of the galaxy  from  the Poisson Equation.

\begin{figure}
\center\includegraphics[width=14.5cm]{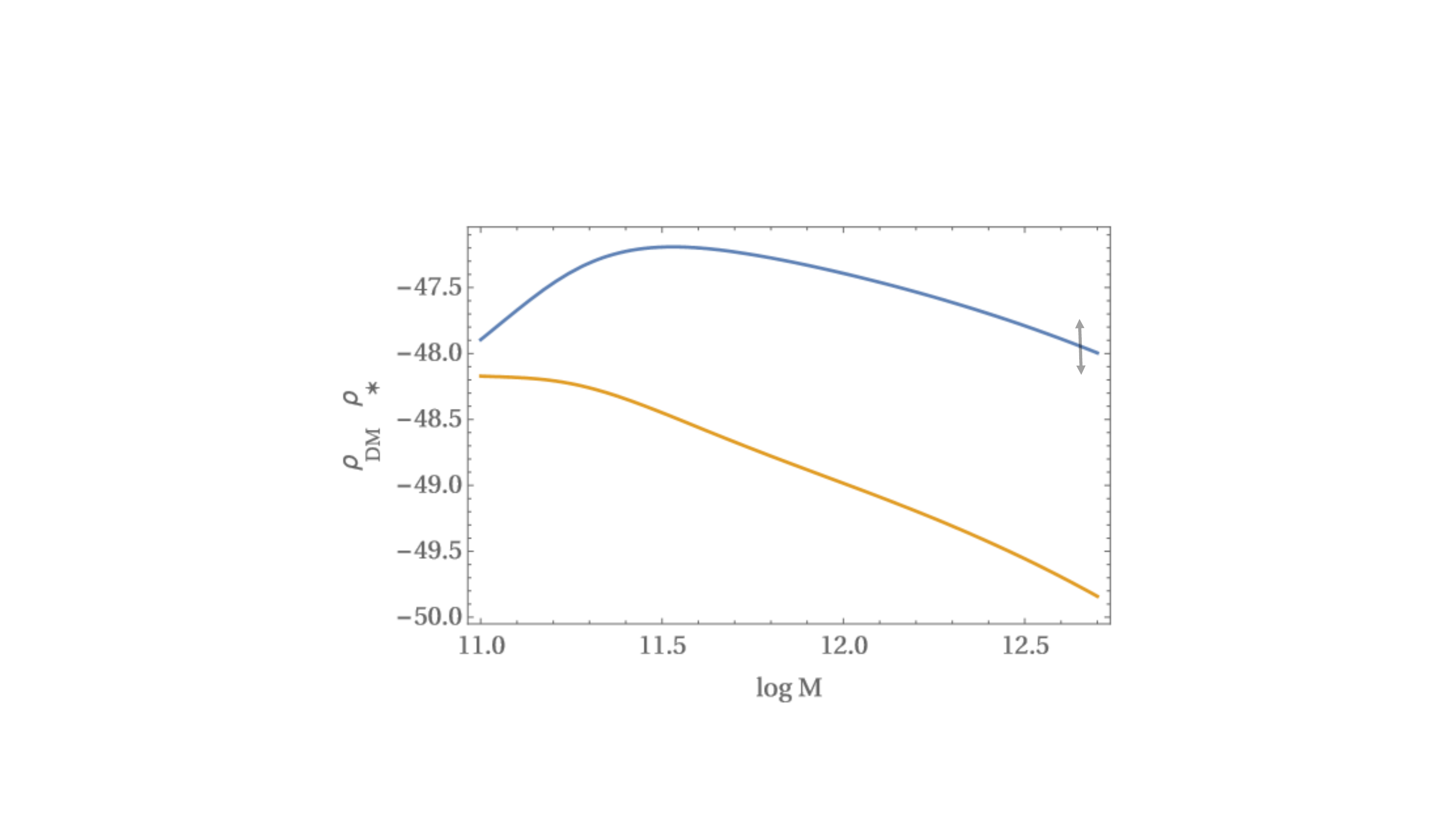}
\caption{ $K_C(R_{cp})$  as function of  the galaxy halo mass $M$.  The 2 $\sigma$ uncertainty is shown with an arrow. The annihilation kernel  at the same radius is also shown (red line) for a comparison}
\end{figure}

 Of course across spirals  $K_C(r)=K_C(r,  M_I)  $,  by varying the galaxy magnitude $M_I$ in    Eqs (4)-(8) of \cite{s7},   we derive $ K_C(r, M_I)$ and so 
 $K_C(R_{cp}(M_I),M_I)$ for the whole family of Spirals.  Remarkably,  we find, see Fig. (5),  that, $K_C(R_{cp}) \simeq  \ const =10^{-47.5} g^2 cm^{-6} $,  within a factor of about 2.  As a comparison,  at the same radius,  $K_{SA}(R_{cp})$ varies by two order of magnitude  with the  galaxy halo mass.  This result in no way is affected  by  the uncertainties on log  $K_C(R_{cp}$  derived  from fitting the URC). In fact, these  are small ($< 0.2 dex $)  and , above all,  independent of  halo mass.

Therefore, in each object, the radius  $R_{cp}$ emerge as an intriguing  lenghtscale  which  1)  marks the radius of  a remarkable feature in the DM pseudo-pressure  distribution 2) it is related to the core radii $r_0$ ,  3) it is almost constant  in all galaxies.  It is impressing  to realize in Fig. 6 that  $ K_C(r,M_h))$   varies hugely among galaxies and at different radiuses and only at $R_{cp}$ it takes a similar value for all objects indicating a sort of a threshold value for the densities in order to have a significant interaction between the two components.

 The two relationships above  may well call for  a general  time  evolution  of the DM density of spirals  $\rho_{DM}(r, t, M_H)$
triggered by a collisional dark-luminous component interaction,  proportional to $K_C$   In this scenario, inside  $\simeq R_{cp}$ of the size of the core radius $r_0$,   the value of the product of the two  densities $K_C $ is  larger than the above  reference value and  enough collisional interactions  have occurred  over a time of $\simeq$ 10 Gyr,  to  flatten the  DM  density distribution.  Instead, outside  $\simeq R_{cp}$,   $K_C$ decreases rapidly with radius  and  collisions get soon  suppressed in any object. 

\begin{figure}
\center
\includegraphics[width=11.5cm]{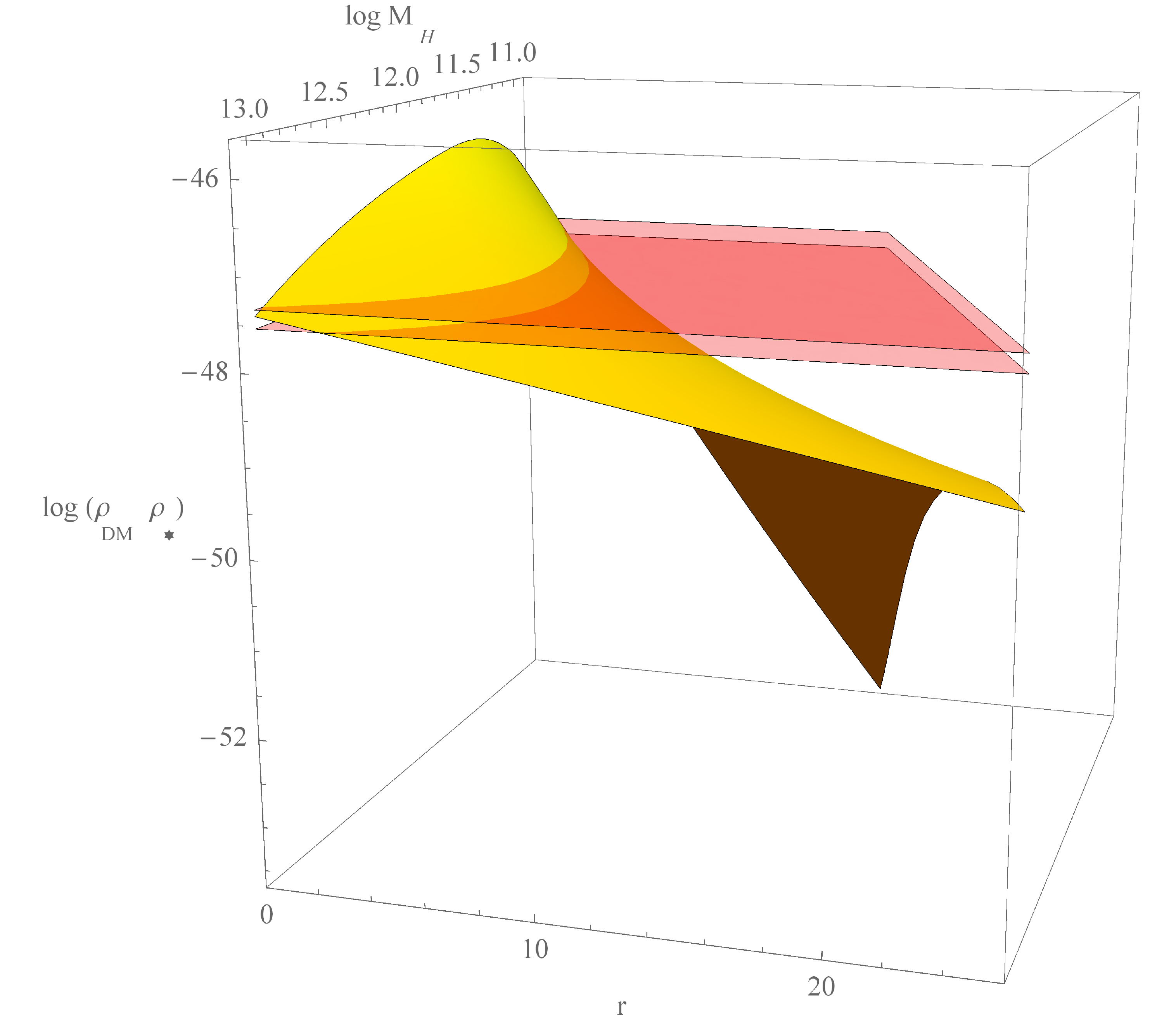}
\caption{ $K_C$  as function of   $log \  (M_H/10^{11} M_\odot)$  and  radius $r/kpc$. The full range of  the values  plus/minus  their 1-$\sigma$  uncertainties of $\sim 0.2$ dex   of  the quantity $K_C(R_{cp}$  lays between the two parallel planes}
\end{figure}

\section{Formations of  cores  in the DM  density }

In the collisional DM scenario,  such kind of  interaction, may  play an important role in shaping   the structural   properties of the Spiral mass distribution  and therefore in the creation of  DM  density cores. This idea also finds support by the fact that    $r_0$ is  related to $R_{cp}$, the lenght-scale of the `collisional' process (see Fig. 4)  and  to $R_{D}($see Fig. 1),  the length scales  of the  stellar distribution.  

The next step is to derive, galaxy by galaxy,   how much dark mass has been involved in the process, i.e.  how many dark particles have interacted with ordinary matter particles. Let us  guess the original density  distribution  of the DM halos, formed in a free fall time of  about  $10^{7-8}$ years , well {\it  before}  that the collisional  interactions, in a time scale of  $10^{10}$ years,  removed a great fraction   of the dark  mass inside $R_{cp}$.   

Noticeably,  the   DM halo  density around Spirals,,  i.e.  outside the region inside which  the collisional  interactions took place, is well reproduced by a  NFW profile with  concentration:  $c= 13 (M/(10^{12} M_\odot))^{-0.13}$  (see \cite{s7}),  so that  in Spirals,  for  $r>R_{cp}$ we have:
 \begin{equation}
\rho_{cusp}(M, r)= \frac{1.65 \times  10^{-24}\times M^{0.073}}{r\left(\frac{k_1 } { M^{0.46}} r +1\right)^2 \left(\log\left(\frac{k_2}{M^{0.13}}+1\right)-\frac{k_2}{\left(\frac{k_2}{M^{0.13}}+1\right) \ M^{0.13}}\right)}
  \end{equation}
with   $k_1=1.82 \times 10^4$ and d  $k_2=4.72 \times 10^2$ and $M$ the halo mass in solar masses. 
Notice that, in  this work,  we  consider  the  latter  distribution  just as an empirical one, survival of the LM-DM interaction occurred at smaller radii.    Extrapolating this density back in time  and to the center of the galaxies, we   recover $\rho_{cusp}$ the  originally  (cuspy) profile  of the just formed  DM halos. In Fig (7) we show the primordial and the  actual  DM density profiles.

 \begin{figure}
\center\includegraphics[width=8.5cm]{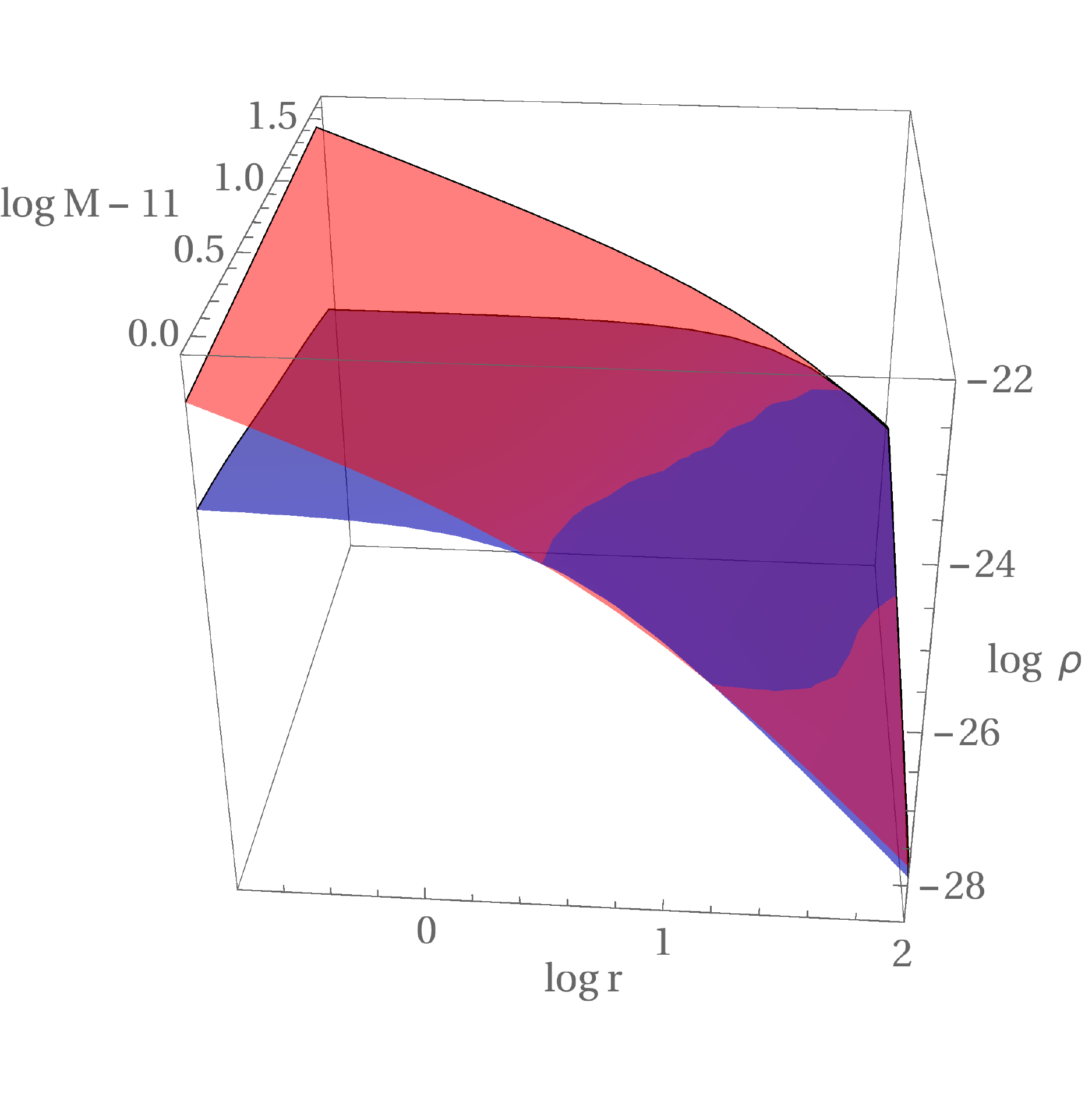}
\caption{Primodia (red) and present-day (blue) DM density profile around galaxies as a function of radius (in kpc) and halo mass (in units of  $10^{11} \ M_\odot $})
\end{figure}
 For a present day spiral of mass $M_H$  the amount of  removed DM over the Hubble time is: 
 
\begin{equation} 
\Delta M_H= 4 \pi  \int^{R_{cp}}_0  (\rho_{cusp}(r, M_H) -\rho(r,M_H)) r^2 dr
 \end{equation}

  The results are in fig (9). We see that in a galaxy of halo mass $10^{11}  M_\odot\leq M \leq 10^{13 M_\odot} $,   the  amount of mass removed by the  dark-luminous collisional interactions is   from $40\  \% $ to $ 90\ \%$,  the original dark matter mass inside the cored radius.

 \begin{figure}
\center\includegraphics[width=9.5cm]{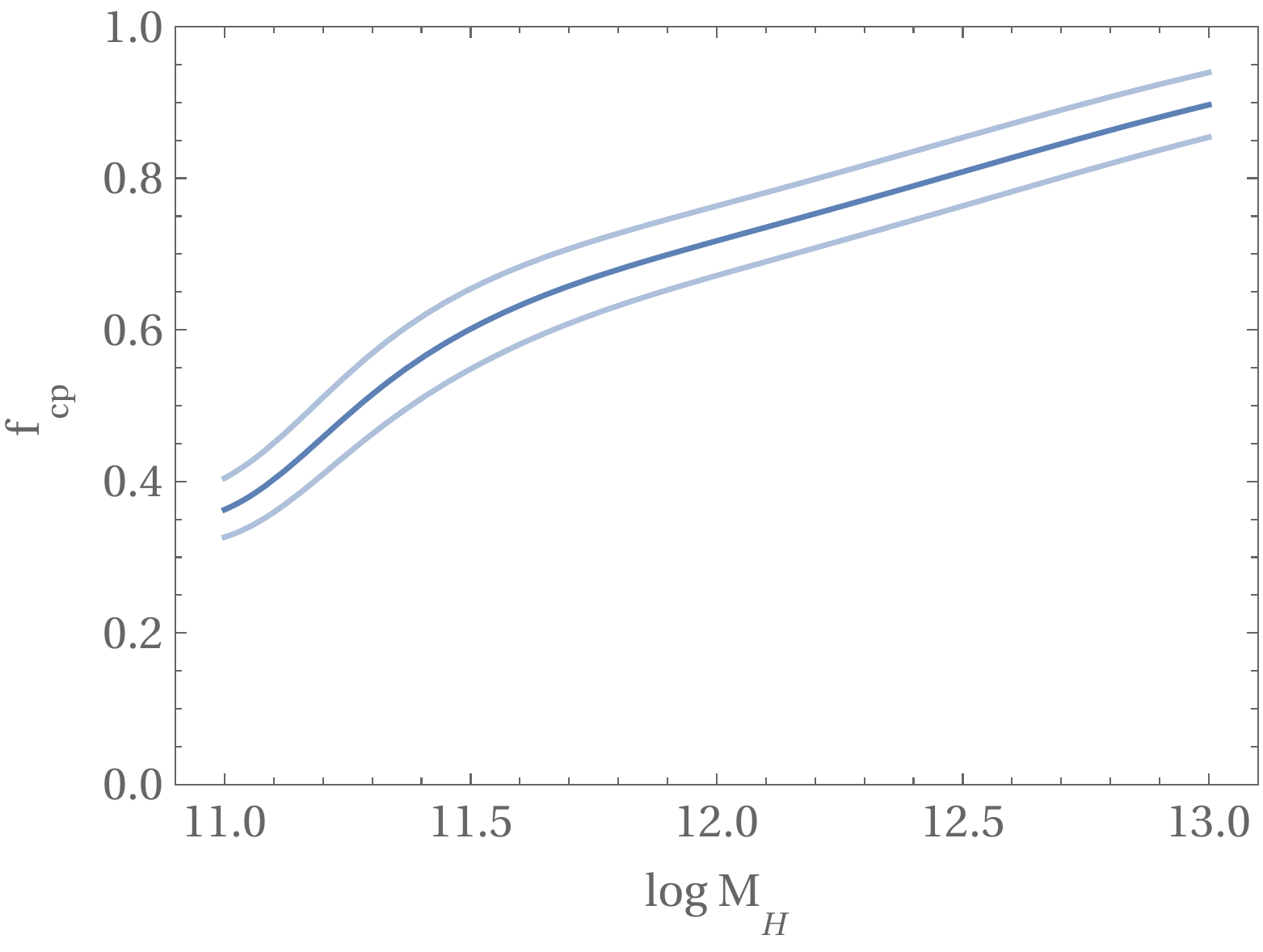} 
\caption{ Removed -to -primordial  dark mass  inside $R_{cp}$, as function of halo mass (blu line). The thinner lines indicate the 1-sigma uncertainty propagated by  those  of the  URC mass model }
\end{figure}
 
 \begin{figure}
\center\includegraphics[width=9.5cm]{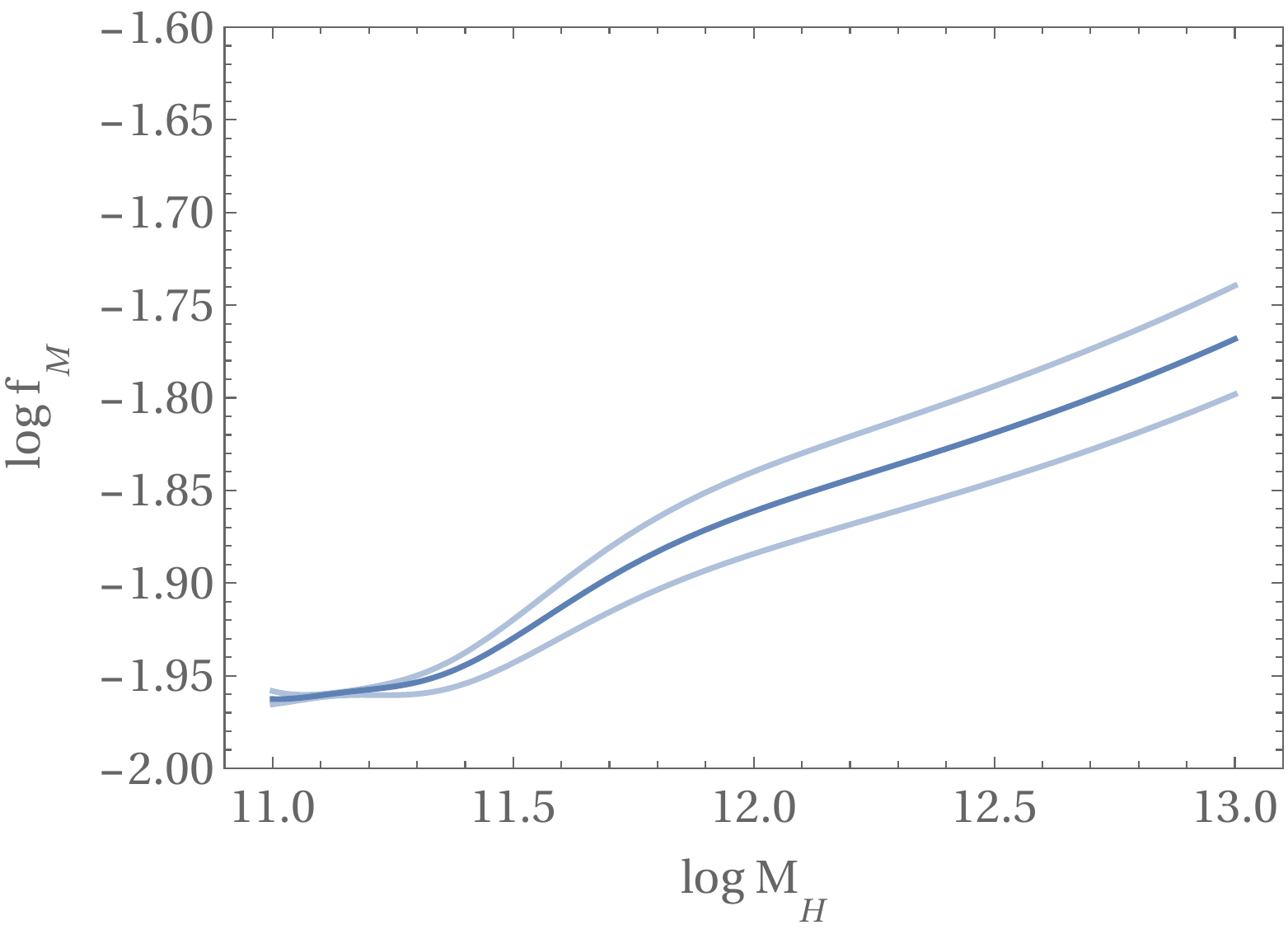} 
\caption{ Removed -to -primordial  dark mass, as function of present  halo mass (blu line). The thinner lines indicate  the 1-sigma uncertainty propagated by  those  of the  URC mass model }
\end{figure}
 However, let us notice that such mass,  likely removed  during the  formation of  the cores in the DM density,   is only $1/100$ the present day halo mass. see Fig (9)
 
 All the intriguing  features in the mass distribution in spirals  could be  created with the complicity of a minuscule fraction of the whole  dark halo mass. This  is in  contradiction with the  current view  according to  which,  each  particle in the  halo  DM has participates  to the formation of the  galaxy   through the   processes of bottom-up collapse and  merging.
 
 \section {Collisional DM and DM halos at z=1}

We have direct support for an evolution of  DM halos  over the  Hubble time as result of  the interactive  process we have claimed for.  We know the structural properties of four disc  galaxies  located at $z=1$ ( \cite {Sz}) from  high spatial resolution  measurements  due  to large boosts  in their apparent   angular sizes that are caused by  strong gravitational lensing from foreground massive galaxy clusters. This provided us with proper  photometry and  kinematics of spirals at  cosmological distances, well outside the  possibility of direct measurements.  The stellar  surface photometry of these galaxies shows that they,   at those early times, had already grown Freeman stellar disks of sizes not much different from those  of local galaxies with the same $V_{opt}$. Their   RCs are  consistent with  the local ones and we  modeled  them   by means of a  Freeman disk and Burkert halo, whose  free parameters, the core radius, the central DM  density and the disk mass  are  derived by standard  best-fitting method. \cite {Sz}.   We find  that the core radii  $r_0$ of the 4 objects are  about $1/3$ those of local spirals with the same $V_{opt}$.  Specifying, at $z=1$,  galaxies have a core radius three time smaller that those  of  the  $z=0$  objects  with  the same   size and  angular momentum  per unit mass of local ones (see Fig. 6).   These 4  galaxies must be formed at $z_f \sim 2-3$ , so the ratio between their age at $z=1$ and  at $z=0$ results to be  $1/3$, a value  which we claim that  it is not a coincidence.

This result, although found in a yet  limited sample, indicates a linear increase  of the  size of the  core radii with time,  a characteristic 
 of secular processes like the one we have proposed.    In the meantime,  it   puts  strong  constraints on  the  idea that the creation of DM density cores  may be  related to supernovae explosions: these  have an exponential evolution with time so that  DM density cores  created by baryonic feedback should already have been  almost fully developed at $z=1$. 

\begin{figure}
\center\includegraphics[width=8.5cm]{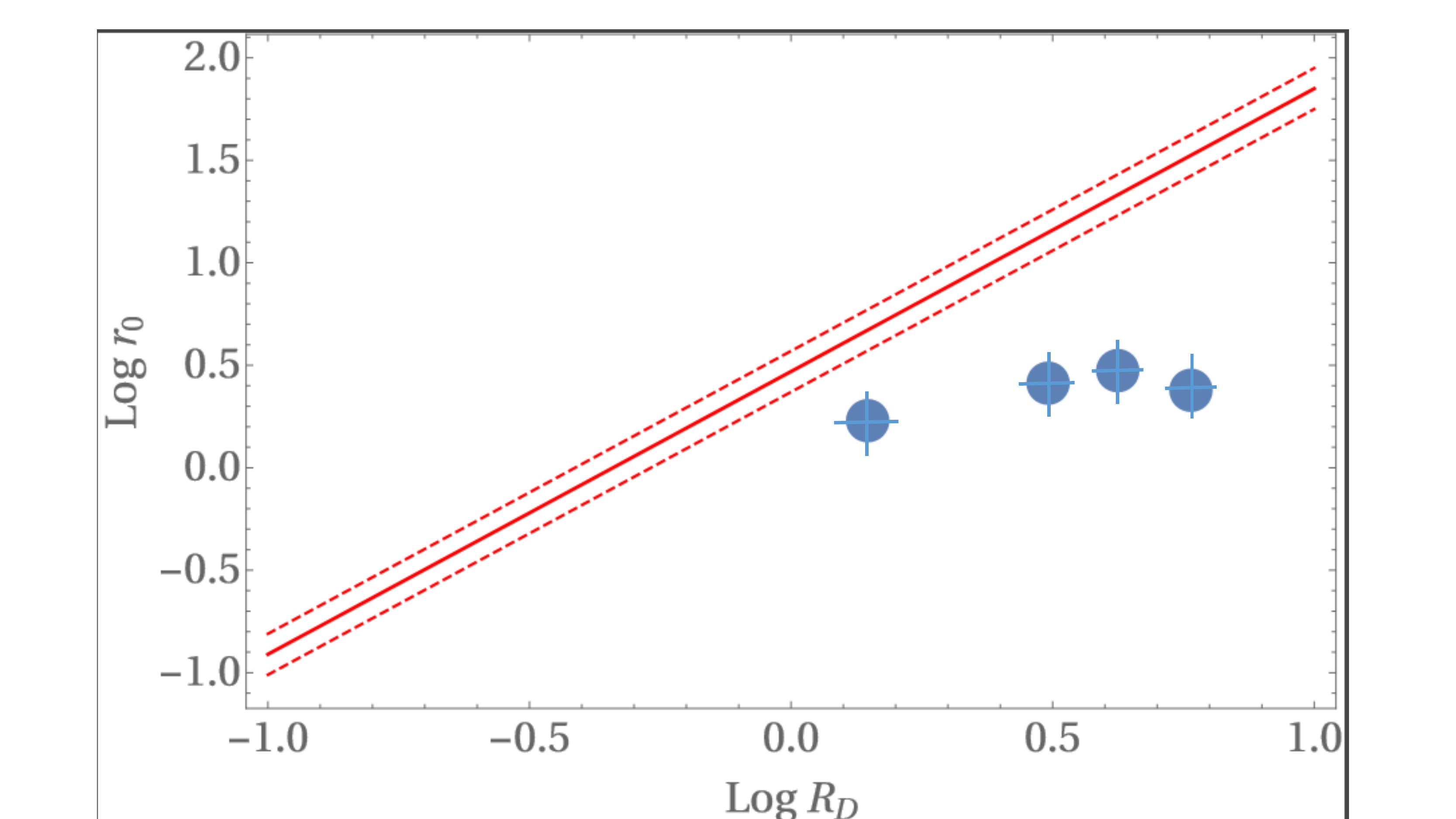}
\caption{The relationship $r_0$-$R_D$ today (lines)  and that for objects at  $z=1$ (points).}
\end{figure}

\section {The cross section of the collisional DM -LM interactions}

We are now in a completely new territory of physics, and,  while  it is relatively easy to point out the existence of new phenomena, it is instead  very difficult to frame them in a  proper  scenario. We have  given evidence that  collisional interactions  may have  been  occurred in Spirals. Given all this,  the kernel of the  collisional interactions  is the product  $\rho(r) \  \rho_*(r)$.  Notwithstanding  the evidence of this interaction,  its cross section can be  estimated only  very roughly. All the action occurs within $R_{cp}$. The DM  is removed  from this region, e.g.  by  absorption (and emission) onto baryonic matter  . DM particles traverses  the region of size $2 R_{cp}$  in a  time twice  the free fall time $t_{ff}= 2.1 \times  10^3 [ <\rho >/(g/cm^{3})]^{-1/2} \  s $.  Now,  we select a galaxy with $M=10^{12} \ M_\odot$, then from  \cite{s7} and from the  previous section:  $R_{cp}=6 \  kpc$, $M_\star(R_{cp})= 4 \times 10^{10} M_\odot$, $  <\rho_\star >=2.9\times 10^{-24} g/cm^3 $ ,  $  <\rho>= 4/3 \ \pi \ M_{cusp}
(R_{cp})/R_{cp}^3= 1.4 \times 10^{-24} \ g/cm^3$ . The number density of the absorbers  inside $R_{cp} $ is $  <n>_\star= 3 /(\pi \  4)  M_\star(R_{cp})/(R_{cp}^3  \ m_H)$ where we have assumed that stars are entirely  made by  hydrogen. The flux attenuation of DM particles, over $10^{10}\  y$,  is 0.5 (see Fig (5)). The number of cycles is $10^{10}/(2 t_{ff})= 10^{10}/{ 10^8}\sim 100$. 

Putting together all these data we roughly estimate the  absorption cross section of  DM vs LM interaction as: : 
 \begin{equation}
 \sigma \sim 10^{-42}\  cm^{-2} \ m_{GeV}
\end{equation}

\section{Conclusion}
 
\begin{figure}
\center\includegraphics[width=8.5cm]{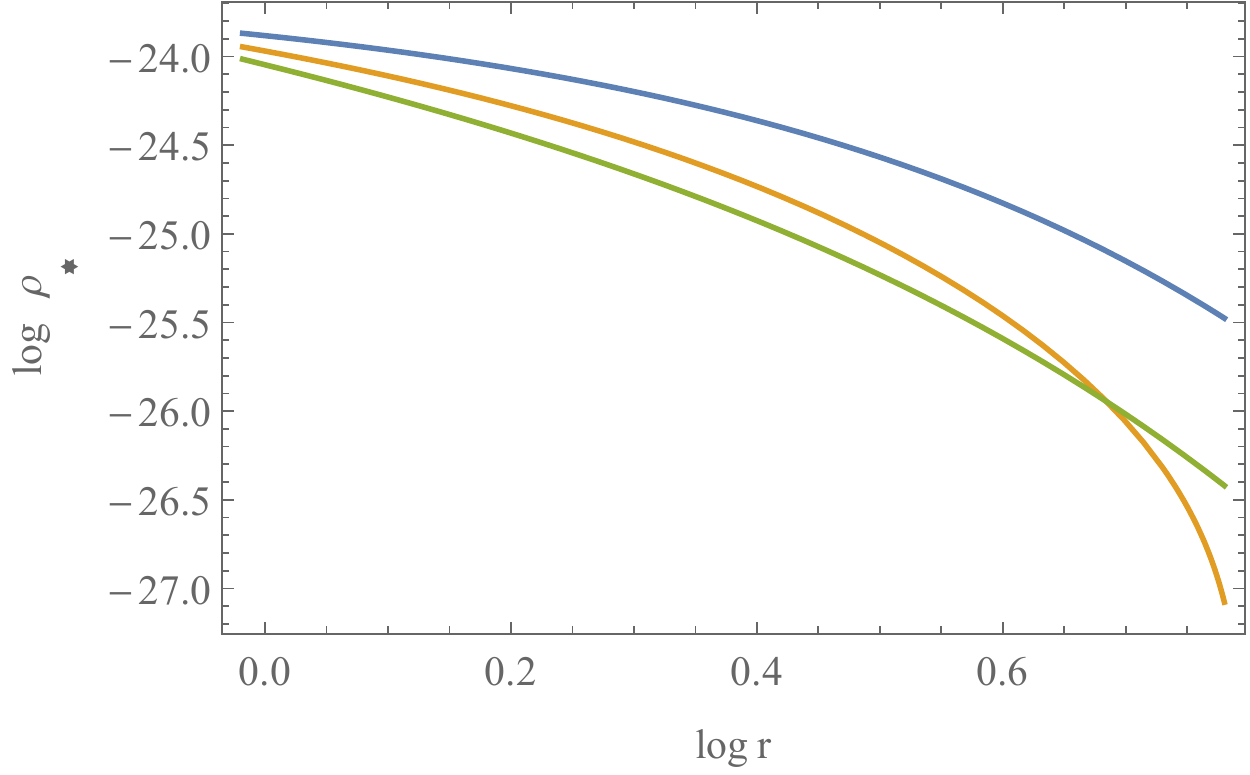} 
\caption{The  various  effective spherical densities for Freeman disks we investigated     ($M_D = 6.3 \times 10^{10} \ M_\odot$,  $R_D= 3.6\  kpc $)}
\end{figure}

Rotation curves studies on a large sample of spiral and dwarf galaxies show an extraordinary correlation between Dark and Luminous matter over many order of magnitude in halo masses. The idea of a dynamical evolution of galaxies, in the Hubble time period, driven only by gravity is failing  the explanation of such a deep-rooted correlation. Moreover, the large discrepancy of DM densities in the inner part of the galaxies from the outcome of NBody $\Lambda$CDM simulations  and the actual  measured data cannot be explained only by astrophysical phenomena without a fine tuned modelization, that very likely could not account for the mounting scenario of  universal correlations. Warm Dark Matter models are certainly  nearer to the latter,  however they  cannot account easily of the universality of its internal structure.
In this paper we also show that the quantity $\rho(r) \rho_*(r)$ tend to be almost the same on all the galaxies as the DM pseudo pressure reaches the maximum close to the Core Radius. The same Pressure has the same value no matter the galaxy dimension. This density product in fact is proportional to the interaction probability of the the two components, Luminous and Dark matter, and account for a direct interaction between them.
This is hardly a coincidence, in that, the  quantity like $K_{SA}= \rho_{DM}^2(r) $ which is proportional to the self interaction of the DM component is varying all over in galaxies and among galaxies 

Therefore, we claim that the structure of the inner parts of the galaxies is driven by a direct interaction between Dark and Luminous components. The DM central cusp, foreseen for any heavy  collisionless DM dark matter particle and also in many other cases, with an increase of DM pressure at lower radius,  gets,  as time goes by, progressively eaten up/absorbed  by the dominant luminous component. The interaction flattens the density of DM and drops the pressure towards the center of the galaxy.

SuSy, if ever appears, has a large energy scale that could in principle lead to neutralino masses in the TeV region. We estimated an absorption cross section for a heavy DM particle from the luminous component to be from 3 to 5 order of magnitude larger than foreseen by SuSy prediction. Current prediction for a minimal Dark Sector zoology based on a (U1) symmetry, with heavy “Dark Fermions” and a mediator “Dark Photon”,
(e.g.\cite{ma,Ri, al,HN})  cannot address the full phenomenology described in this paper. A direct coupling between Dark Fermions and SM particles, in presence of luminous matter, should take place allowing the decay of the heavy Dark Matter particles into light ones, SM or other. Heavy Axion Like Particles (ALP) are another DM candidate.  Finally There can be  in principle also be mediators between the two sectors, such as the Higgs.

The estimated absorption cross section is low  enough to make direct DM detection experiments based on nucleus recoil fail, the recoil could not happen at all and the experiments should see large energy showers, that if they are initiated by electrons or photons can be confused with high energy neutrino showers. On the contrary,  exotic searches at LHC, with the integrated luminosity already taken and foreseen in the next years, can confirm such picture. Production cross section at high energy should be big enough to have the DM particles produced in detectable quantity. Moreover, the positron excess in our galaxy, if from DM interaction, points to a TeV scale mass particle that might be related with our proposal.
LHC searches are not calibrated to detect such type of invisible particles, the method based on missing transverse energy and momentum asymmetry, is not very sensitive if the production cross section is of the femtobarn order of magnitude. The large QCD background is masking all the events. Missing mass detection done with more sophisticated apparatus such as the forward spectrometers CT-PPS and AFP, respectively in CMS and Atlas, can enhance the detection probability if the production is initiated by photon-photon interaction. Also triggering algorithms can be optimized for such a search, right now they are focused on SuSy production and exotics searches are done with low efficiency.

The DM scenario that arises from this picture points to a DM sector different from  any predicted so far.  Neutralino interaction cross sections foreseen by SuSy cannot account for the absorption rate we measure. So a new DMP or a more complex DM sector should appear. We yet don't have any hint on how the coupling with matter is, surely nothing yet envisaged in the present DM theoretical panorama. “Model independent” searches have to be pursued in LHC. While in direct searches underground the detection could be made looking to the appearance of particle showers in large mass detectors as neutrino telescopes (Ice Cube, Antares...). A diffuse gamma ray signal with large energy, above 100GeV, with a cutoff spectrum, could be correlated to DMP absorption.

\section{Acknowledgments} We tank the referee for important help in presenting the results of this paper.

\section{Appendix A}

Other used 3D  modelization of the stellar disk intrinsically  2D distribution different from that adopted in this paper are: i)  to make spherical the stellar disk mass profile. This, in cylindrical coordinates is given, for the Freeman disk, by$ M_D(r)=M_D (1-(1+r/R_D) e^ {-r/R_D}))$ the 3D density is obtained by substituting the cylindrical  coordinate $R$ with the spherical $r$ so that: $\rho_\star(r)=
1/(4 \pi \ r^2) dM_D(r)/dr $, ii) By solving the Poisson equation one finds that a Freeman disk  mimics  a sphere of effective density  $\rho_\star(r)=GM_D/R_D^3 H(r/R_D)$ where the latter  function is given in Eq.  (9) of \cite{sal10}. 

\section{Appendix B}
\begin{figure}
\center\includegraphics[width=16.5cm]{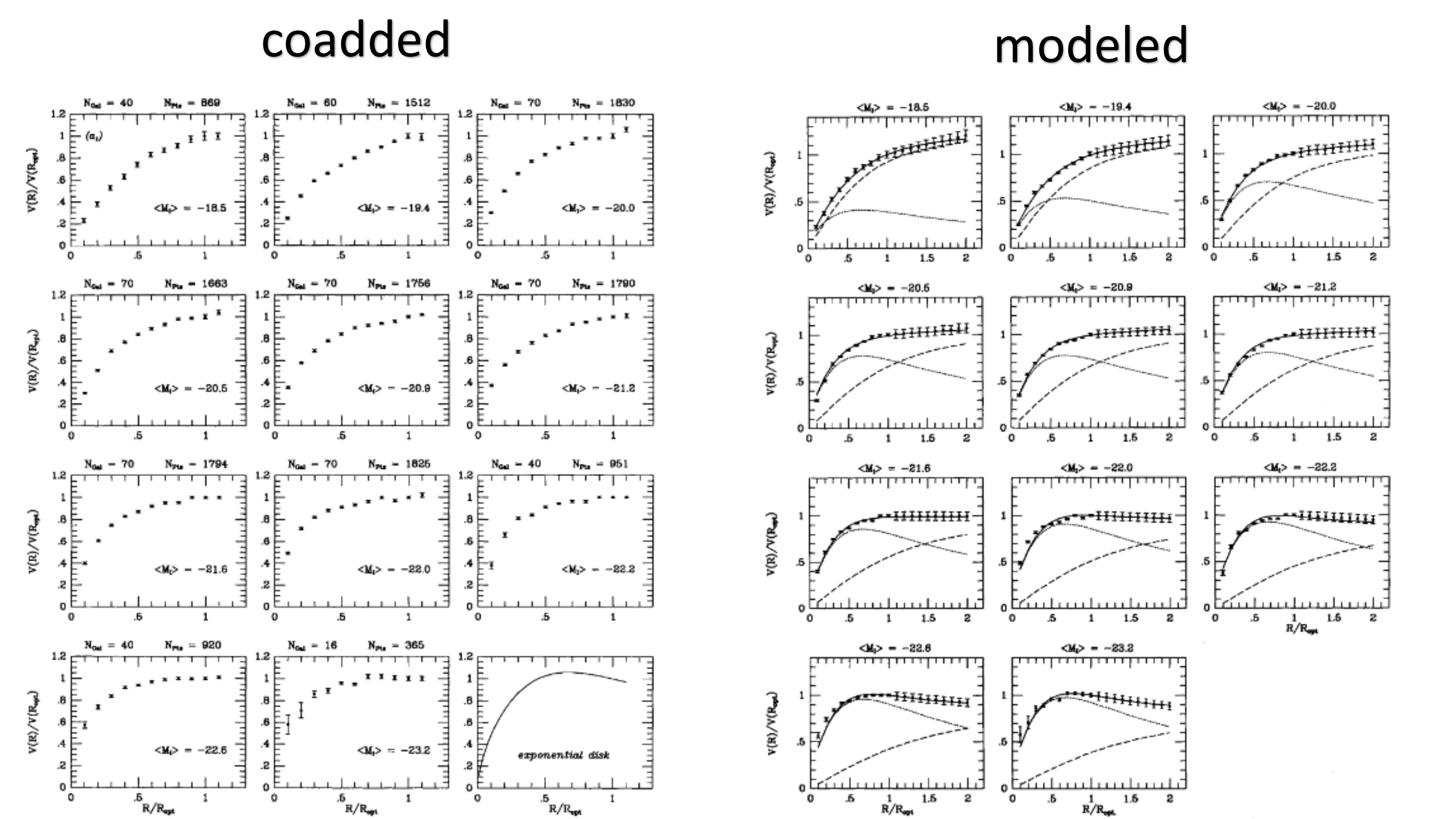} 
\caption{Number of objects and data,  the 11 coadded RCs and their velocity modelling}
\end{figure}
We  show, by means of fig  a) the very great number of objects and of  individual velocity measurements that have concurred in the derivation of the 11 coadded rotation curves representing  the whole kinematics of spirals. b) the smoothness and the very small internal rms  $\delta v/v =0.02 - 0.05$ of these latter c) the goodness of the DM halo + disk  best fit velocity  model to data, which leads  to the small  fitting uncertainties of Eq(4) and in turn, to the existence of the URC in Spirals

\vfill\eject
 
\end{document}